\documentclass[sigconf, nonacm]{acmart}

\usepackage[utf8]{inputenc}
\usepackage{geometry}
\usepackage{algorithm}
\usepackage{algpseudocode}

\usepackage{balance}
\usepackage{subcaption}
\usepackage{enumitem}

\newcommand{\ipso}{I1S$^4$o}

\begin{document}
\title{Defeating duplicates: A re-design of the LearnedSort algorithm}

\author{Ani Kristo}
\affiliation{%
  \institution{Brown University}
}
\email{ani@brown.edu}

\author{Kapil Vaidya}
\affiliation{%
  \institution{MIT}
}
\email{kapilv@mit.edu}

\author{Tim Kraska}
\affiliation{%
  \institution{MIT}
}
\email{kraska@mit.edu}

\begin{abstract}
LearnedSort is a novel sorting algorithm that, unlike traditional methods, uses fast ML models to boost the sorting speed. The models learn to estimate the input's distribution and arrange the keys in sorted order by predicting their empirical cumulative distribution function (eCDF) values. 

LearnedSort has shown outstanding performance compared to state-of-the-art sorting algorithms on several datasets, both synthetic and real. However, given the nature of the eCDF model, its performance is affected in the cases when the input data contains a large number of repeated keys (i.e., duplicates).

This work analyzes this scenario in depth and introduces LearnedSort 2.0: a re-design of the algorithm that addresses this issue and enables the algorithm to maintain the leading edge even for high-duplicate inputs. Our extensive benchmarks on a large set of diverse datasets demonstrate that the new design performs at much higher sorting rates than the original version: an average of 4.78$\times$ improvement for high-duplicate datasets, and 1.60$\times$ for low-duplicate datasets while taking the lead among sorting algorithms.  

Source code and benchmarks can be found at \url{https://github.com/learnedsystems/LearnedSort}
\end{abstract}

\maketitle

\section{Introduction}
LearnedSort is a novel sorting algorithm that, unlike traditional methods, uses fast ML models to boost the sorting speed. We introduced the algorithm in 2020\cite{kristo:sigmod2020} together with a large set of benchmarks that showed outstanding performance as compared to state-of-the-art sorting algorithms. LearnedSort uses a model to estimate the input's distribution and arranges the keys in sorted order by predicting their empirical CDF (eCDF) values. While the idea behind distribution sorts is not new, LearnedSort was the first algorithm to take advantage of learned models and achieve extremely fast sorting rates. 

Nevertheless, given the nature of eCDF modeling, LearnedSort's performance was affected when the input data contains a large number of repeated keys. This results from the model making the same eCDF predictions for equal keys, hence causing a large number of key collisions and bucket overflows that led to a performance slowdown. Interestingly, recent evaluations from other sorting algorithm papers\cite{isp4o_benchmark, blacher2021fast} focus entirely on \textit{synthetic} datasets to determine the ``best'' algorithm. While it is arguable whether those datasets are representative of real-world scenarios or indicative of the performance of an algorithm in practice, real datasets can unquestionably contain an arbitrary number of duplicate keys. Hence, in order to make LearnedSort more practical, it is important to make it more robust against duplicates, as the data distribution and the number of duplicates are not known \textit{a priori}. 

This work shows a detailed description of what triggers this degradation and introduces LearnedSort 2.0: a re-design of the algorithm that enables it to maintain the leading edge even for high-duplicate inputs. We also present extensive benchmarks to evaluate its performance on a set of diverse datasets with various sizes and distribution properties. In particular, we benchmark on several real datasets that contain extremely high ratios of duplicate keys (upwards of 95\%). In all the experiments, the new design performs at much higher sorting rates than the original version: an average of 4.78$\times$ improvement for high-duplicate datasets, and 1.60$\times$ for low-duplicate datasets in which the original LearnedSort was already the most competitive algorithm.

The following section gives a background on the original LearnedSort algorithm, summarizing its high-level design and highlighting the experimental results presented in \cite{kristo:sigmod2020}. Next, we describe why its performance suffered for inputs containing a large portion of duplicate keys and continue with strategies for addressing this issue. In Section 4, we introduce LearnedSort 2.0, describing its architecture in detail, followed by comprehensive benchmarks in Section 5.

\section{Background on Learned Sort}

\begin{figure*}
    \centering
    \includegraphics[width=\textwidth]{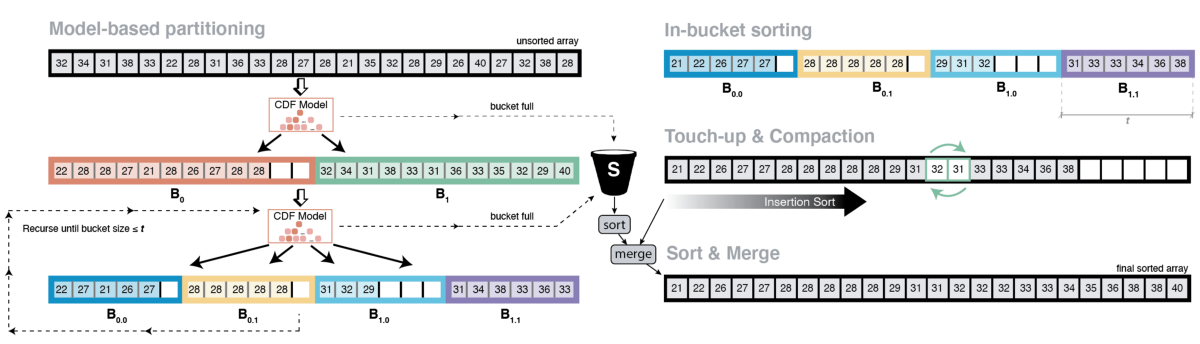}
    \caption{The original design of LearnedSort. The first step is the recursive partitioning of the keys into buckets based on the eCDF model predictions. Once the keys have been fully partitioned, all buckets are internally sorted using a model-based Counting Sort. Lastly, small prediction errors are fixed with an Insertion Sort pass, and the elements in the spill bucket are merged with the final output.}
    \label{fig:original_diag}
\end{figure*}

The core idea of the LearnedSort algorithm was simple: we used ML techniques to train a model that estimated the distribution of the data from a sample of the keys; then used it to predict the order of keys in the sorted output. Our learned model takes a key (x) and returns the count of keys in the input that are less than x. Therefore, the model acts as an estimator of the scaled eCDF of the input.


Despite the simplicity of the core idea, the reality is that there existed multiple challenges with this approach, such as making sure that the distribution modeling was extremely fast and accurate, and that the algorithm leveraged the cache effectively to avoid any performance degradations.

We looked at these challenges in-depth in \cite{kristo:sigmod2020} and addressed them for the design of LearnedSort: We adjusted the model so that it partitioned input keys into fixed-capacity buckets based on the predicted eCDF values. The process continued for another round until the buckets get small. At that point, the algorithm stopped the partitioning process and used a model-based, Counting Sort-style subroutine to sort the elements inside the buckets, followed by a clean-up step for minor imperfections in the sorted order. When these fixed-capacity buckets got full, it used a "spill bucket" that collected all the overflowing keys. The spill bucket was sorted separately using std::sort (an optimized Quicksort hybrid) and merged with the other keys at the very last step. Fig. \ref{fig:original_diag} illustrates the key steps of the LearnedSort algorithm.

Eventually, this design of LearnedSort resulted in a very fast, cache-efficient, ML-enhanced sorting algorithm that showed impressive results in our benchmarks. LearnedSort achieved an average of 30\% better performance than the next-best sorting algorithm (\ipso), 49\% over Radix Sort, and an impressive 238\% better than std::sort- the default sorting algorithm in C++\cite{kristo:sigmod2020}.

\section{The Achilles' heel of Learned Sort}
\begin{figure}
    \centering
    \includegraphics[width=\linewidth]{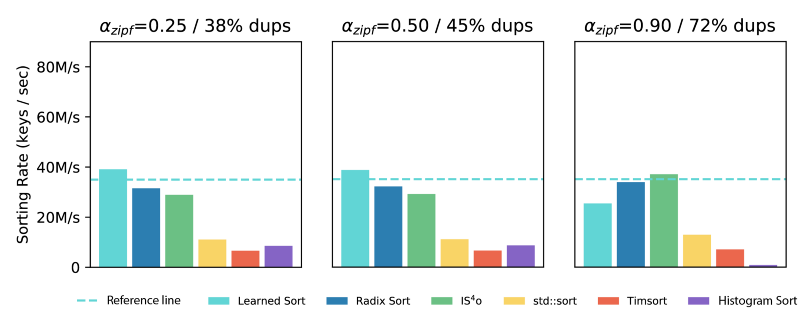}
    \caption{The sorting rate of LearnedSort on synthetic Zipfian datasets with increasingly more duplicate keys (Higher is better). The reference line shows the sorting rate of LearnedSort when the data has no duplicates. LearnedSort's performance deteriorates after the duplicate proportion reaches 50\%.\cite{kristo:sigmod2020}}
    \label{fig:original_zipf}
    \vspace{-10pt}
\end{figure}

One of the most important challenges we had to address was to minimize the impact of too many duplicate keys. In such scenarios, the eCDF 
values for any two equal keys in the input array A would also be equal ($\forall x_1, x_2 \in A,  x_1 = x_2 \implies eCDF(x_1) = eCDF(x_2)$). Therefore, the learned model makes the same predictions and places them onto the same bucket. The algorithm could tolerate a certain degree of duplicate keys; however, this became an issue when the input contained a substantial amount of such keys (usually more than 50\%). In this case, certain buckets would quickly reach full capacity and overflow onto the spill bucket, while others would be mostly empty. In turn, since the size of the spill bucket increased, progressively more keys needed to be sorted using an external algorithm (i.e., std::sort, which is a slower algorithm). In addition, all of these keys would need to be merged back at the final step of the algorithm (see Fig. \ref{fig:original_diag}). Thus, the spill bucket sorting step became the bottleneck, and the performance of LearnedSort on high-duplicate inputs deteriorated compared to the average case. Fig.\ref{fig:original_zipf} shows that LearnedSort's sorting rate decreases by 22\% in the case when the input data had a Zipfian distribution with skew 0.9, which corresponds to 72\% duplicates.

LearnedSort could tolerate a certain degree of duplicate keys by using an ``exception-based" approach: The model would detect duplicate keys at training time from the input sample and use an associative data structure that counted how many times these "problematic" keys occurred. When it encountered such keys at sorting time, it would update their occurrence counts and skip the partitioning steps. At the final stage, the algorithm re-wrote these keys back onto the sorted output based on the tracked counts. This strategy helped avoid some bucket overflows and consequently the inflation of the spill bucket size.

However, this approach is not optimal since, for each incoming key, it requires a table look-up to determine if it is a problematic key. This look-up happens in the main loop, therefore increasing the complexity of this critical section. In order to have constant-time look-ups, we used a hash table; however, this unfortunately required additional subroutines for table sorting and merging with the output. Nevertheless, regardless of the goodness of constant-time look-ups, hashing operations are relatively expensive for performance-sensitive routines like sorting.

Therefore, it was necessary for LearnedSort to have a comprehensive solution that could handle any degree of duplicates while avoiding expensive data accounting operations at sorting time.

\begin{algorithm}
\caption{LearnedSort 2.0 partitioning algorithm}\label{alg:lsv2:part}
 \scriptsize
\begin{flushleft}
    \hspace*{\algorithmicindent} \textbf{Input} $A$ - the input keys\\
    \hspace*{\algorithmicindent} \textbf{Input} $F_A$ - the CDF model for the distribution of $A$\\
    \hspace*{\algorithmicindent} \textbf{Input} $p$ - the partitioning level (0 or 1)\\
    \hspace*{\algorithmicindent} \textbf{Input} $f$ - bucket fanout\\
    \hspace*{\algorithmicindent} \textbf{Input} $c$ - bucket fragment capacity\\
    \hspace*{\algorithmicindent} \textbf{Output} $S_B$ - bucket sizes\\
    \hspace*{\algorithmicindent} \textbf{Output} $S_F$ - fragment sizes\\
\end{flushleft}
\begin{algorithmic}[1]
\Function{Partition}{$A, F_A, p, f, c$}
    \State $Frag\gets [][]$
    \Comment{An array of $f$ fragments of capacity $c$}
    \State $S_B\gets [0]\times f$
    \Comment{Records bucket sizes}
    \State $S_F\gets []$
    \Comment{Records fragment sizes}
    \State $w\gets 0$
    \Comment{Write head for fragments}
    \item[]
    \For{$x\in A$} 
        \State $pos\gets F_A(x) \cdot f^{1+a} - p(i\cdot f)$
        \Comment{Predict the bucket index}
        \State{$Frag[pos].append(x)$}
        \Comment{Place in the respective bucket fragment}
        \State{$S_B[pos]++$}
        \Comment{Update the bucket size}
        \If {$Frag[pos].size == c$}
        \Comment{Fragment is full}
            \State{Copy $Frag[pos][0..c]$ onto $A[w]$}
            \Comment{Copy back the fragment}
            \State{$w \gets w + c$}
            \Comment{Update the write head}
            \State{$S_F$.append($c$)}
            \Comment{Record fragment size}
            \State{$Frag[pos].clear()$}
            \Comment{Reset the content of the fragment}
        \EndIf
    \EndFor
    \item[]
    \State{\textit{// Copy the remaining fragments that were not full}}
    \For{$i \in [0, f]$}
        \State{$s \gets Frag[i].size$}
        \State{Copy $Frag[i][0..s]$ onto $A[w]$}
        \State{$S_F$.append($s$)}
        \State{$w \gets w + s$}
        \State{$Frag[pos].clear()$}
    \EndFor
    \State{\textbf{return $S_B, S_F$}}
\EndFunction
\end{algorithmic}
\end{algorithm}

\begin{algorithm}
\caption{LearnedSort 2.0}\label{alg:lsv2}
 \scriptsize
\begin{flushleft}
    \hspace*{\algorithmicindent} \textbf{Input} $A$ - the input to be sorted\\
    \hspace*{\algorithmicindent} \textbf{Input} $F_A$ - the eCDF model for the distribution of $A$\\
    \hspace*{\algorithmicindent} \textbf{Input} $f$ - number of buckets (fan-out)\\
    \hspace*{\algorithmicindent} \textbf{Input} $c$ - bucket fragment capacity\\
\end{flushleft}
\begin{algorithmic}[1]
\Function{LearnedSort}{$A, F_A, f, c$}
    \State{\textit{// Model-based partitioning}}
    \State{$S_B, S_F \gets \textsc{Partition}(A, F_A, 0, f, c)$} 
    \Comment{Returns bucket sizes and fragment sizes}
    \item[]
    \State{\textit{// Place fragments belonging to the same bucket contiguously}}
    \State{$\textsc{Defragment}(A, S_B, S_F)$}
    \item[]
    \State{\textit{// Re-partition each bucket in $f$ sub-buckets}}
    \State{$r \gets 0$} 
    \Comment{Read-head}
    \For{$i \in [0, f]$}
        \State{$n \gets S_B[i]$}
        \Comment{$n$ is the size of the current bucket}
        \State{$s \gets r$}
        \Comment{Bucket start index}
        \State{$e \gets s + n$}
        \Comment{Bucket end index}
        \If{\textsc{Is-Homogeneous}($A[s..e]$)}
            \State{$r\gets r + n$}
            \Comment{Skip this bucket and update the read-head}
        \Else{}
            \State{$S^\prime_B, S^\prime_F \gets \textsc{Partition}(A[s..e], F_A, 1, f, c)$}
            \State{$\textsc{Defragment}(A[s..e], S^\prime_B, S^\prime_F)$}
            \Comment{Defragment the sub-buckets}
            \item[]
            \State{\textit{// Iterate through the sub-buckets}}
            \For{$j \in [0, f]$}
                \State{$n^\prime \gets S^\prime_B[j]$}
                \Comment{$n$ is the size of the current sub-bucket}
                \State{$s^\prime \gets r$}
                \Comment{Sub-bucket start index}
                \State{$e^\prime \gets s^\prime + n^\prime$}
                \Comment{Sub-bucket end index}
                \If{$\neg\textsc{Is-Homogeneous}(A[s^\prime..e^\prime])$}
                    \State{\textit{// Model-based Counting Sort}}
                    \State {$K\gets [0] \times n^\prime$}
                    \Comment{Counting histogram}
                    \For{$k\in [0..n^\prime]$}
                        \State{$adj \gets (i\cdot f + j)  / f^2$}
                        \Comment{Prediction adjustment offset}
                        \State{$pos\gets \left[F_A(A[s^\prime+k]) - adj\right] \cdot A.size $} 
                        \Comment{Predicted position in array}
                        \State{$K[pos]++$}
                        \Comment{Update histogram count}
                    \EndFor
                    \item[]
                    \For{$k \in [1..n^\prime]$}
                    \Comment{Calculate histogram running totals}
                        \State{$K[k]\gets K[k]+K[k-1]$} 
                    \EndFor
                    \item[]
                   \State{$T\gets[]$}
                   \Comment{Temporary auxiliary memory}
                    \For{$k \in [0..n^\prime]$}
                    \Comment{Order keys w.r.t. the cumulative counts}
                        \State{$adj \gets (i\cdot f + j)  / f^2$}
                        \State{$pos\gets \left[F_A(A[s^\prime+k]) - adj\right] \cdot A.size $}
                        \State{$T[K[pos]\gets A[s^\prime+k]$}
                        \State{$K[pos]--$}
                    \EndFor 
                    \State{Copy $T[0..n^\prime]$ back to $A[s^\prime]$}
                \EndIf
                \State{$r\gets r + n^\prime$}
                \Comment{Update the read head}
            \EndFor
        \EndIf
    \EndFor 
    \item[]
    \State{\textit{// Touch-up for minor model prediction errors}}
    \State{$\textsc{Insertion-Sort}(A)$}
\EndFunction
\end{algorithmic}
\end{algorithm}

\section{Learned Sort 2.0}

LearnedSort was the fastest algorithm for almost all of the datasets that we evaluated it on, including synthetic distributions (Uniform, Normal, Exponential, Lognormal, Multimodal, Zipfian, and TPC-H), and real ones from OpenStreetMaps, Facebook Graph, and WebLogs\cite{kristo:sigmod2020}. However, the common thing about all these datasets was that they contained almost no duplicate keys. Therefore, on average, less than 5\% of the input elements would go onto the spill bucket, making it a non-critical segment of the algorithm. 

\subsection{A candidate strategy: Predicting problematic buckets}
A natural evolution of the key-flagging approach in the original algorithm is to use the learned model to predict and flag buckets that might overflow exceedingly. These buckets would be the ones containing the highly repeated keys and, consequently, be responsible for the enlargement of the spill bucket. For these buckets, instead of performing the re-partitioning and counting-sort procedures that would normally follow, LearnedSort could switch to a third-party deterministic sorting algorithm that is highly efficient on high-duplicate scenarios. (Examples of such algorithms are usually variations of Quicksort that use equality buckets. In this case, the keys are partitioned three ways: less-than, equal-to, and greater-than the selected pivot value(s). For example, \ipso\cite{isp4o_benchmark} is one of the most efficient sorting algorithms that benefits from such cases.) The caveat for this approach is that the model is required to assign larger capacities to the flagged buckets to prevent duplicate keys from landing onto the spill bucket again. For this, the model could use an inexpensive counting histogram at training time based on the small sample collected from the input. 

This strategy worked much better than the original one since it avoided the performance costs associated with the use of additional data structures for logging problematic keys. Now, all we needed to use was a simple bitset that indicates problematic buckets. However, in most of the real datasets that we benchmarked, the flagged buckets appeared to contain more than 90\% of the keys--sometimes even up to 99\%, meaning that LearnedSort would depend on an external sorting algorithm for a very large portion of the input. In the best case, for high-duplicate inputs, LearnedSort would be no faster than the external algorithm used, in addition to the overhead from the model training and partitioning procedures. On the other hand, the performance on low-duplicate inputs was not affected. Therefore, given our goal for designing a highly efficient LearnedSort, we had to consider more radical design changes.

\subsection{The winning strategy: Partitioning in-place and eliminating the spill bucket} 

\begin{figure}	
	\centering
	\begin{subfigure}[t]{\linewidth}
		\centering
		\includegraphics[width=\linewidth]{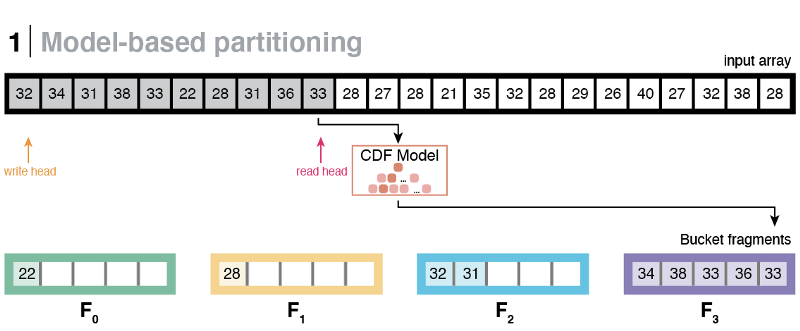}
		\caption{Step 1 of the LearnedSort 2.0 algorithm. The keys are placed in bucket fragments based on the eCDF model's predictions.}\label{fig:step1}		
	\end{subfigure}
	\quad
	\begin{subfigure}[t]{\linewidth}
		\centering
		\includegraphics[width=\linewidth]{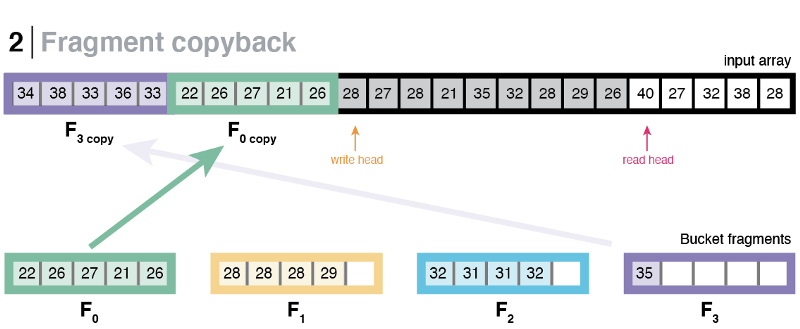}
		\caption{Once fragments get full (i.e., $F_3$ and $F_0$), they are copied back to the input array and their contents are cleared.}\label{fig:step2}
	\end{subfigure}
	\begin{subfigure}[t]{\linewidth}
		\centering
		\includegraphics[width=\linewidth]{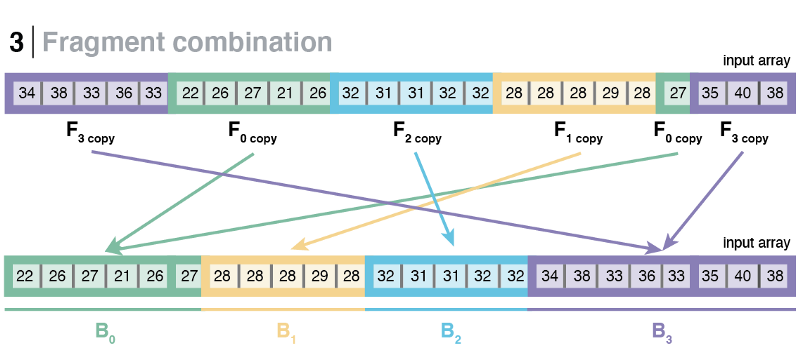}
		\caption{Fragments belonging to the same buckets are combined so that they form a contiguous space. After this step, bucket boundaries are well defined.}\label{fig:step3}
	\end{subfigure}
	\begin{subfigure}[t]{\linewidth}
		\centering
		\includegraphics[width=\linewidth]{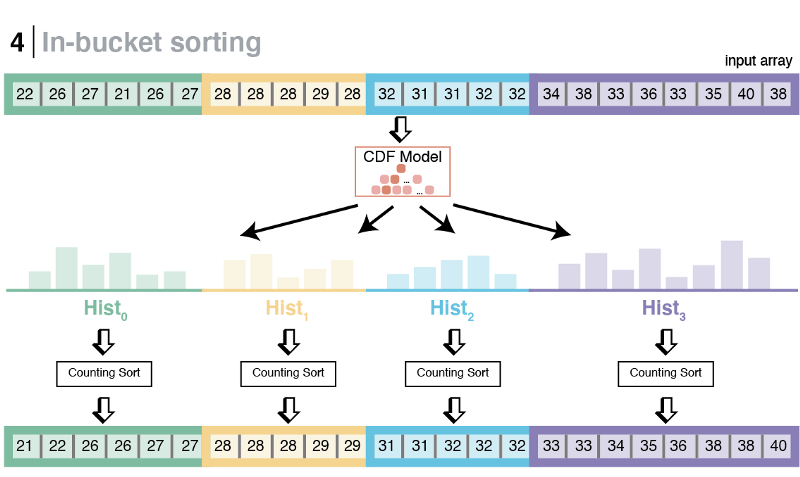}
		\caption{After the second partitioning, the keys inside each bucket are sorted using a model-based Counting Sort subroutine.}\label{fig:step4}
	\end{subfigure}
	\caption{The design of LearnedSort 2.0}\label{fig:new_diag}
\end{figure}

It was clear that to improve LearnedSort's performance on high-duplicate inputs, we had to eliminate the spill bucket, and the possibility of bucket overflows. To achieve this, we could not allocate the same capacity to every bucket anymore; on the contrary, we had to allocate precisely as much space as each bucket needed. One approach is to do a linear scan over the input and, for each key, predict the bucket index and increment its corresponding size counter. However, the overhead for this additional step is not insignificant. To mitigate that cost, we could use the training sample for estimating each bucket's size before allocating. In that case, however, we have to make room for estimation errors by over-allocating, and additional memory acquisition also results in a slowdown.

Instead, what worked best was to think of buckets as a collection of smaller, fixed-capacity fragments. The eCDF prediction step would be the same; however, instead of placing the key directly to its predicted bucket, we place it to the corresponding pre-allocated fragment owned by the predicted bucket (Fig. \ref{fig:step1}). Once a fragment reaches its full capacity (i.e., 100 elements), it is copied back to the input at a given write-head (Fig. \ref{fig:step2}). This operation overwrites the input's content; however, it is guaranteed that the overwritten elements have already been scanned and placed onto bucket fragments. The write-head is updated, and the copied fragment is cleared to make space for more incoming elements. This procedure continues until the entire input has been processed. 

Using this method, we managed to avoid using a spill bucket altogether and logically give each bucket as many slots that they needed, but at the cost of bucket fragmentation. After the first partitioning step has finished, the fragments owned by the same bucket (sibling fragments) will most probably be not contiguous, rather scattered throughout the input array. Therefore, it is necessary to perform an additional step that combines sibling fragments to form a contiguous space that represents a whole bucket  (Fig. \ref{fig:step3}). This is simply the procedure that is also used in file defragmentation utilities in most Windows and Linux operating systems\cite{linux_defrag,windows_defrag}.

After the buckets have been defragmented, the algorithm will again re-partition the keys inside each bucket to further refine the sortedness of the input. This step is similar to the original LearnedSort, and it is done in two steps as a way to maximize data locality and cache utilization. While iterating through the formed buckets, the new algorithm performs a quick check to see if all the elements in the bucket are equal, in which case, it may skip the re-partitioning and jump to the next bucket.

At the end of the second partitioning step, the keys will be quasi-sorted in that, for any bucket $B_i$, all the keys $x_i \in B_i$ and $x_{i+1} \in B_{i+1}$,  $x_i \leq x_{i+1}$, and so on. However, we don't have any sort of guarantee on the order of the keys \textit{inside} each bucket since they have not been placed in any particular order. Therefore, the next step is to do an in-bucket sort using the same model-enhanced Counting Sort subroutine as in the original LearnedSort\cite{kristo:sigmod2020} (Fig. \ref{fig:step4}). This subroutine is fast and has linear time complexity.

Finally, LearnedSort 2.0 still uses Insertion Sort as a final and deterministic touch-up subroutine, which, in almost linear time, guarantees that the input has been monotonically ordered. Unlike the original LearnedSort, there will not be any more steps involving spill bucket sorting and merging (compare to Fig. \ref{fig:original_diag}). The complete pseudocode for LearnedSort 2.0 and its new partitioning subroutine can be found in Alg. \ref{alg:lsv2} \& \ref{alg:lsv2:part}

\section{Related Work}
There exist numerous sorting algorithms that introduce various novel techniques to achieve good performance, such as leveraging vectorized instructions or branchless conditionals. However, they generally tend to be classified in two groups: \textit{comparison-based} or \textit{distribution-based}. 

\subsection{Comparison sorts}

Even though the most well-known comparison-based sorting algorithm is Quicksort, variations that are practically much faster than a vanilla implementation. Examples of Quicksort-based algorithms are std::sort, BlockQuicksort, and \ipso. 

Std::sort is the default sorting algorithm for C++, provided by the GNU Standard Template Library\cite{std::sort}. It works by combining the partitioning scheme of Quicksort, except that it switches to Insertion Sort when the partition's size goes below a given threshold. When the recursion depth exceeds the logarithm of the input size, it instead switches to Heapsort to avoid the worst-case complexity of Quicksort.

On the other hand, BlockQuicksort\cite{blockquicksort} improves on Quicksort by processing elements in blocks instead of on one-by-one basis. In this way, they aim to decouple the control from data flow and achieve an increase of 80\% in sorting speed with respect to the GCC implementation of std::sort. 

\ipso\cite{isp4o_benchmark} is a newer Quicksort-based sorting algorithm that also uses the batching approach like BlockQuicksort, in addition to using multiple pivot points and an entirely in-place partitioning. They store the pivots in a BST that efficiently leverages conditional instructions to make branchless decisions for which partition each element goes to. This, in turn, leads to much lower branch misprediction rates. In addition, they also use equality partitions besides the regular less-than and greater-than partitions as an optimization for early termination of recursion. 

In their most recent benchmarks, they also include the original LearnedSort algorithm; however, it is important to address that their observations were affected by the presence of a known implementation bug\cite{bug}. In addition, the benchmark did not include any real data, therefore not capturing more complex and interesting data patterns. In this work, we aim to provide a comprehensive set of benchmarks on a diverse set of real and synthetic datasets. Therefore, we avoided using sorted, reverse-sorted, or identical-element datasets since they would be trivially captured by an \textsc{if-else} check at the beginning of the function call.

\subsection{Distribution sorts}
Another big family of sorting algorithms is distribution-based sorts, which generally tend to be faster in practice than the comparison-based ones. 

The most popular distribution-based sort is Radix Sort, which has a linear-time complexity with respect to the input size. This is a generalization of the Counting Sort algorithm that minimizes the size of the counting histogram by making multiple passes over the data. This also indicates that Radix Sort is sensitive to the key length since that would require more passes. Naturally, Radix Sort is mostly suited for integer keys; however it can also be adapted for floating points\cite{radix_sort}. 

SkaSort is another distribution sort based on RadixSort. It provides several optimizations on top of RadixSort regarding generalization for different data types, optimization of the inner loop, and making it in-place. These improvements resulted in $2-3\times$ higher sorting speed than std::sort\cite{skasort}, making SkaSort still a very competitive sorting algorithm.

\section{Benchmarks}

\begin{figure*}
\centering
	\begin{subfigure}[t]{\linewidth}
		\centering
        \includegraphics[width=\linewidth]{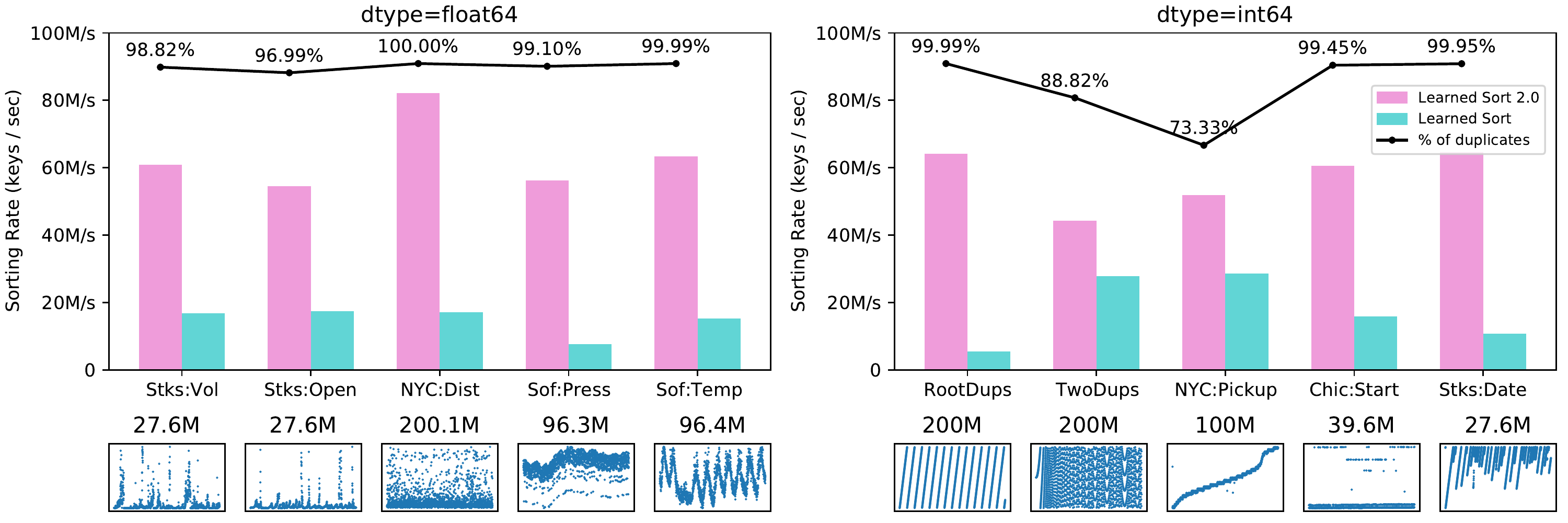}
        \caption{High-duplicate datasets}
        \label{fig:hi_dups}	
	\end{subfigure}
	\quad
	\begin{subfigure}[t]{\linewidth}
		\centering
        \includegraphics[width=\linewidth]{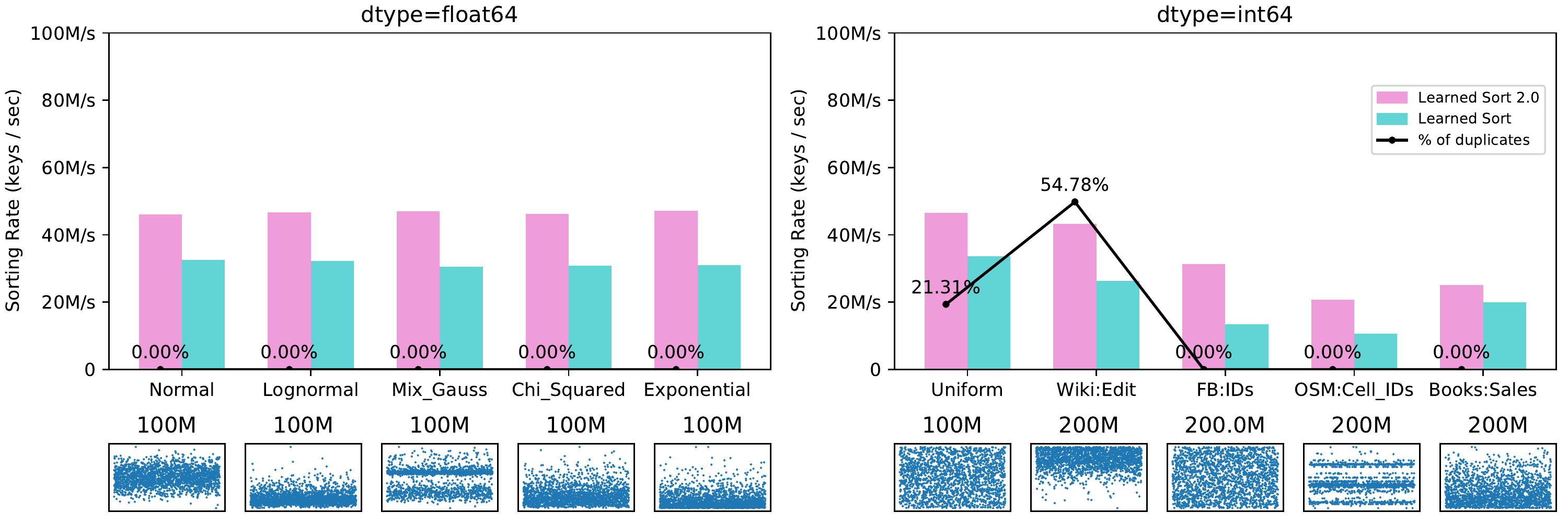}
        \caption{Low-duplicate datasets}
        \label{fig:lo_dups}
	\end{subfigure}
	\caption{The sorting rate comparison of LearnedSort 2.0 and LearnedSort on a mix of real and synthetic datasets (higher is better). The key distribution and size of each dataset are visualized in the plots beneath the performance charts. The ratio of duplicate keys in each dataset is plotted in the red line above the bar charts.}
\end{figure*}

The new design of LearnedSort 2.0 resulted in remarkable performance improvements as compared to the original one. It is on average 4.33$\times$ faster for high-duplicate \textit{real} datasets and 6.57$\times$ faster for high-duplicate \textit{synthetic} datasets. For all other cases, LearnedSort 2.0 receives a 1.60$\times$ performance speed-up from the original algorithm. The following sections describe the experimental setup and analyze the performance gains in more depth.

\begin{figure*}
    \centering
    \includegraphics[width=\linewidth]{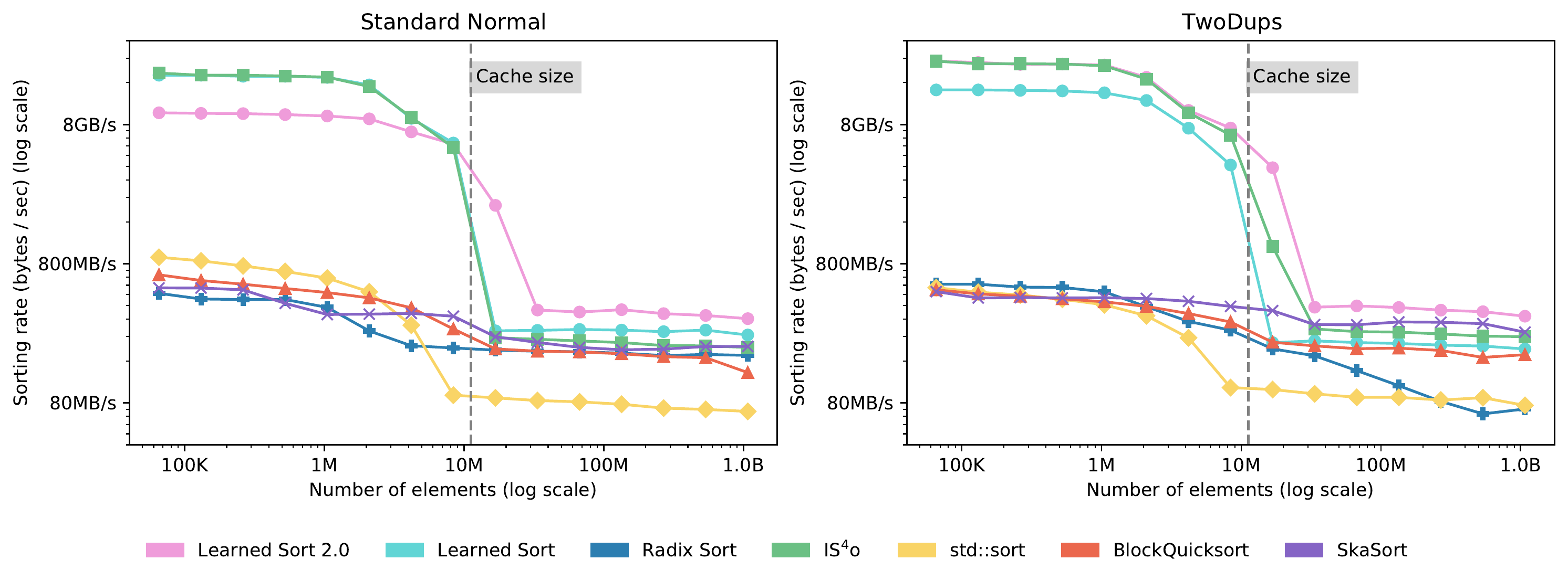}
    \caption{The sorting rate of LearnedSort 2.0 and other competitive sorting algorithms in Standard Normal datasets of increasing input size (1M up to 1B keys). The vertical dashed line indicates the L2 + L3 CPU cache size. Both axes are in logarithmic scales. Higher is better.}
    \label{fig:scalability}
\end{figure*}

\subsection{Setup}
All benchmarks were performed and measured on a server-grade Linux machine running on Intel® Xeon® Gold 6150 CPU @ 2.70 GHz with 376GB of memory, and compiled with GCC 0.2 with the -O3 flag for full optimizations. The model we used for training has two layers and 1000 leaf models, trained with a uniformly selected 1\% sample of the input. The default fan-out parameter was 1000, and the fragment capacities were capped at 100.  

The following is a detailed explanation of some of the real and synthetic datasets used:

\begin{itemize}[topsep=1pt,itemsep=0pt,parsep=0pt,leftmargin=10pt]
    \item \textbf{Stks:[Vol, Open,Date]} -- This dataset contains historical daily opening prices for all tickers currently trading on NASDAQ (stocks and ETFs), trading volumes, and trading dates. The historical data is retrieved from Yahoo finance via yfinance python package. It contains prices for up to 01 of April 2020\cite{stks}.
   
    \item \textbf{NYC:[Dist,Pickup]} -- The yellow taxi trip records include fields capturing pick-up datetimes and trip distances.\cite{nyc}.
   
    \item \textbf{Sof:[Press,Temp]} -- The Sofia dataset contains time-series air quality metrics measured from outdoor sensors in Sofia, Bulgaria. The data to be sorted represents air pressure and temperature at 1-minute intervals\cite{sof}.
    
    \item \textbf{Chic:[Start]} - The Chicago Taxi Trips dataset contains taxi trips reported to the City of Chicago in its role as a regulatory agency in the last six years. The data to be sorted represents the trip starting timestamp \cite{chic}.
    
    \item \textbf{Wiki:[Edit]} -- The Wikipedia dataset contains article edit timestamps \cite{sosd}.
    
    \item \textbf{FB:[IDs}] -- The FB dataset contains an upsampled set of Facebook user IDs from a random walk in the FB graph\cite{sosd}.
    
    \item \textbf{OSM:[Cell\_IDs]} -- The OpenStreetMaps dataset contains uniformly sampled locations IDs from the entire world, represented as Google S2 Cell IDs\cite{sosd}.
    
    \item \textbf{Books:[Sales]} -- The Books dataset contains book sale popularity data from Amazon\cite{sosd}.
    
    \item \textbf{Mix\_Gauss} -- A synthetic random additive distribution of five Gaussian distributions whose means and variances are also chosen at random. The weights of the distributions are also chosen at random.
    
    \item \textbf{RootDups}\cite{blockquicksort} -- This distribution is such that for an array $A$ of size $N$, $A[i] = i \mod \sqrt{N}$.
    
    \item \textbf{TwoDups}\cite{blockquicksort} -- Similar to RootDups, except the formula is $A[i] = i^2 + N / 2 \mod N$.

    \item \textbf{Uniform}($0-N$), \textbf{Normal}($\mu=0, \sigma=1$), \textbf{Lognormal}($\mu=0, \sigma=0.5$), \textbf{Chi\_Square}($k=4$), \textbf{Exponential}($\lambda=2$).

\end{itemize}

\subsection{Performance in High-Duplicate Datasets}

\begin{figure}
    \centering
    \includegraphics[width=\linewidth]{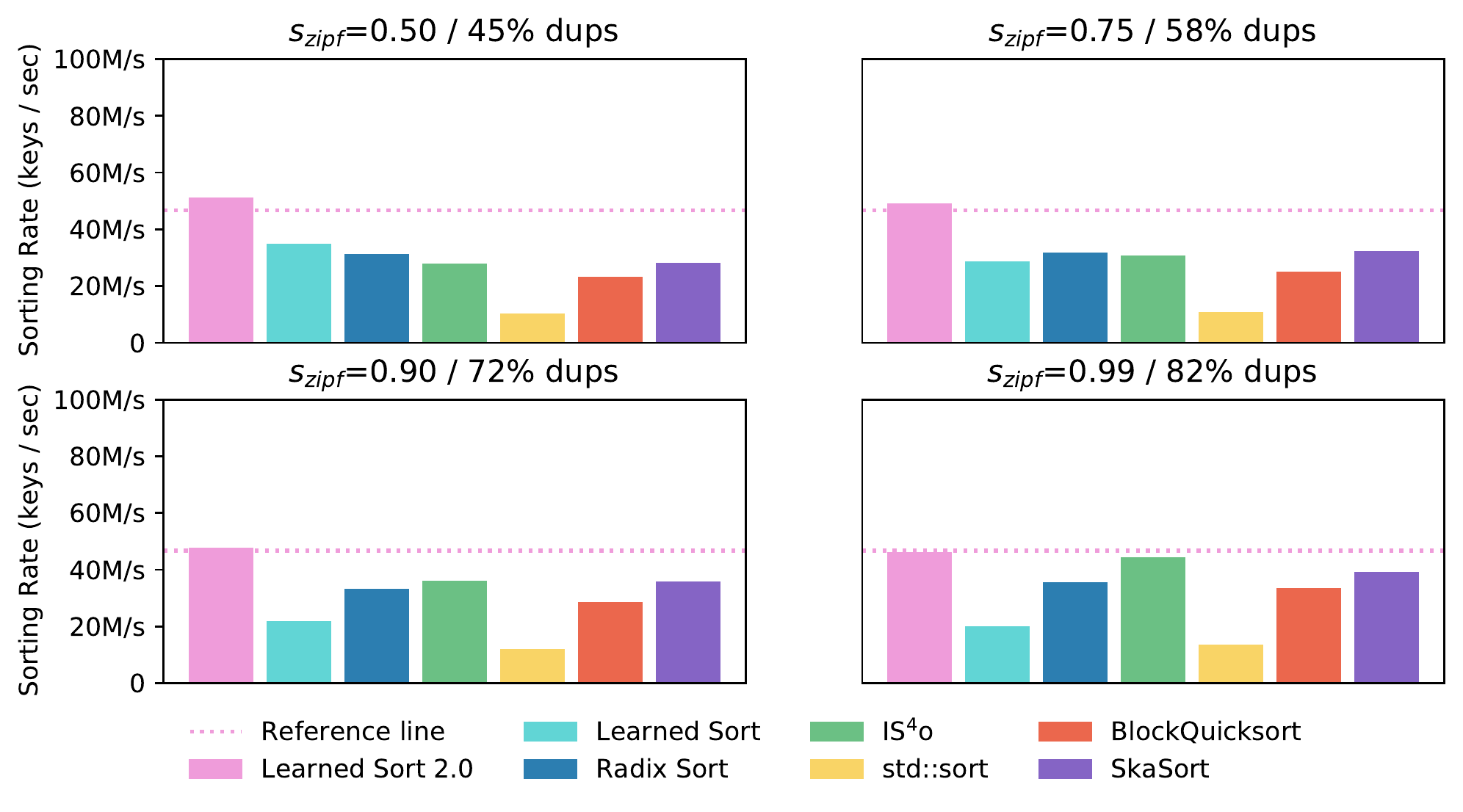}
    \caption{The sorting rate of LearnedSort 2.0 and other state-of-the-art sorting algorithms in Zipfian datasets of an increasing skew parameter (0.50 - 0.99), which corresponds to an increase in the portion of duplicates (shown at the top of the plots). The reference line indicates the sorting rate of LearnedSort 2.0 on a Standard Normal distribution that has no duplicate keys. (Higher is better)}
    \label{fig:zipf}
\end{figure}

We first evaluated LearnedSort 2.0 against the original algorithm on ten different datasets from the list above. Figure \ref{fig:hi_dups} shows the sorting rate of LearnedSort and LearnedSort 2.0 on a mix of real and synthetic datasets whose majority of keys is comprised of duplicates. In all of the cases, LearnedSort 2.0 gets a major performance boost: It is \textit{at least} 60\% faster than the original LearnedSort (in the case of TwoDups), and \textit{on average} 378\% faster for all these datasets.

Next, we show the performance of LearnedSort 2.0 in comparison with other state-of-the-art sorting algorithms on Zipfian datasets with progressively increasing skew--which corresponds to a higher proportion of duplicates. As shown in Figure \ref{fig:zipf}, the sorting rate of LearnedSort 2.0 remains fairly unchanged by the increasing ratio of duplicates, dropping only 3\% below the reference line in the case of the highest skew parameter. At the same time, LearnedSort 2.0 leads with the best performance overall, remaining highly competitive with respect to other algorithms even for datasets with a large degree of duplicates.

\subsection{Performance in Low-Duplicate Datasets}

In order to further analyze the improvements in LearnedSort 2.0, it is also important to see how the algorithm performs on datasets that LearnedSort was already very good at. We have demonstrated  that the original LearnedSort algorithm was the best-performing one in datasets that contained a low degree of duplicates\cite{kristo:sigmod2020}, and this re-design was not aimed at improving in those aspects. Nonetheless,  LearnedSort 2.0 still outperforms the original algorithm by an average of 60\%. Figure \ref{fig:lo_dups} shows this benchmark's results. 

\subsection{Scalability}

We can also look at how LearnedSort 2.0 scales compared to the original version and other algorithms. We compared its sorting rate in inputs drawn from a Standard Normal (0\% dups), and TwoDups distributions (89\% dups) with sizes varying from 1M up to 1B of doubles, and the results are shown in Figure \ref{fig:scalability}. 

In the case of the Standard Normal distribution, LearnedSort 2.0 is on average 34\% faster than the original algorithm when the data size is larger than the cache size, whereas, for smaller inputs, it is up to 45\% slower. However, the running times for such small sizes are less than 5 milliseconds, and the performance difference is only 0.5–1 ms. More notably, LearnedSort 2.0 outperforms the next-best algorithm (\ipso) by an average of 65\%, and Radix Sort by 95\%. 

On the other hand, in the case of TwoDups, LearnedSort 2.0 gets a 56\% improvement over the original algorithm for input sizes smaller than the cache size and 78\% improvement for those larger than the cache. In addition, LearnedSort 2.0 is 29\% faster than the next-best algorithm (SkaSort) and 38\% faster than \ipso.

\subsection{Micro-benchmarks}

\begin{figure}	
	\centering
	\begin{subfigure}[t]{\linewidth}
		\centering
		\includegraphics[width=\linewidth]{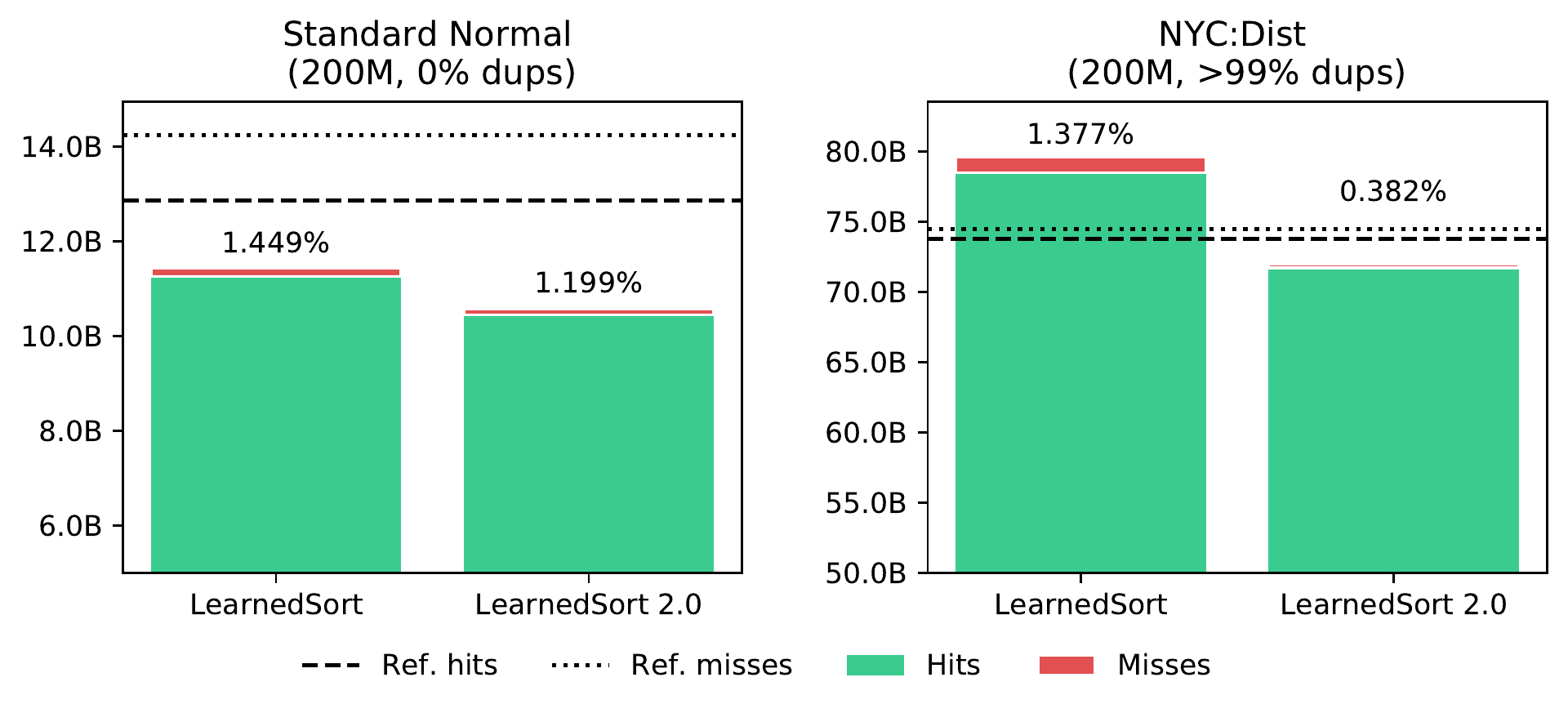}
		\caption{Branch hits and misses (y-axis does not start at 0)}\label{fig:branches}		
	\end{subfigure}
	\quad
	\begin{subfigure}[t]{\linewidth}
		\centering
		\includegraphics[width=\linewidth]{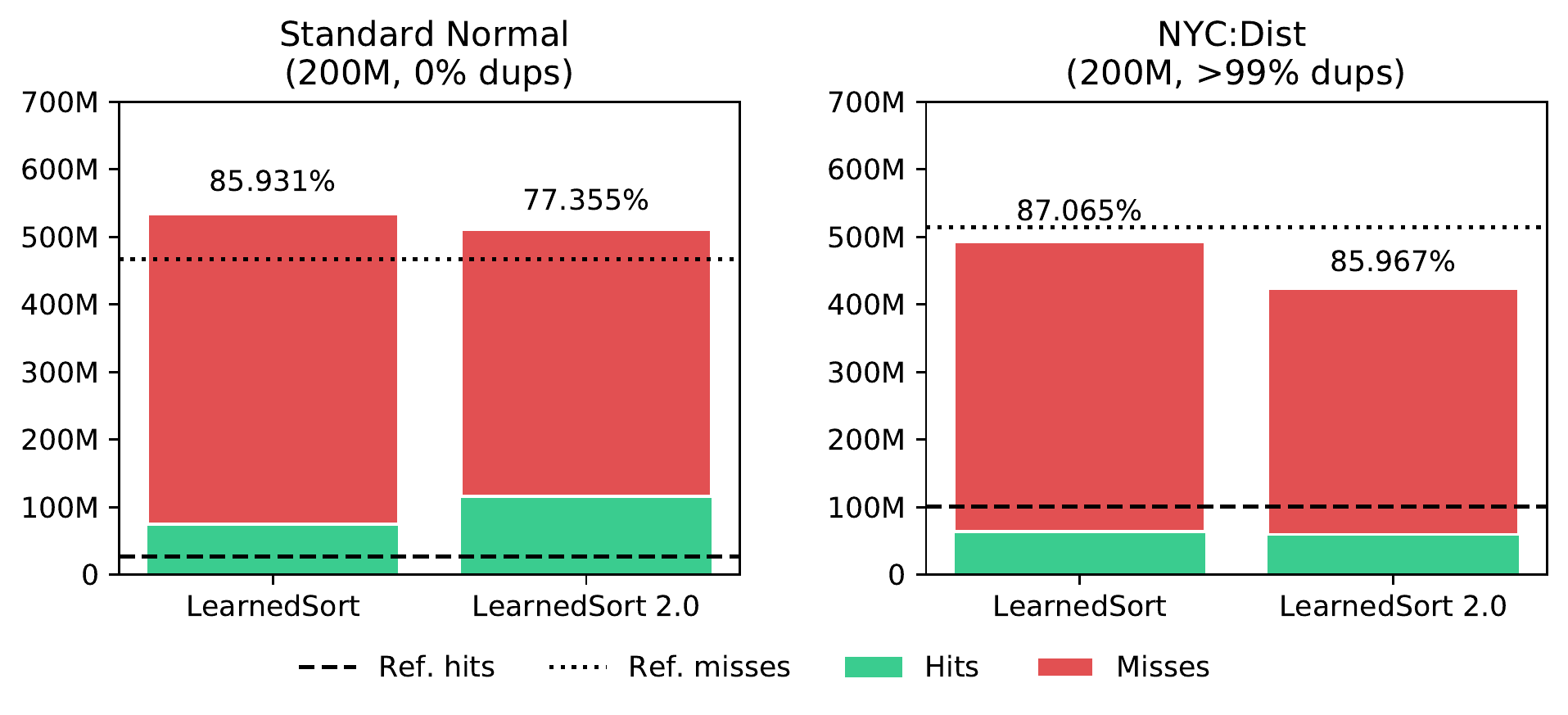}
		\caption{Cache hits and misses}\label{fig:cache}
	\end{subfigure}
	\caption{Comparison of hit-rates and miss-rates for cache references and branches for LearnedSort and LearnedSort 2.0. The dashed reference lines represent the average hit rates from the baseline algorithms, and the dotted lines represent the average miss rates from the other baseline algorithms. The data labels on top of the bar charts show the number of misses over all references. For a clearer visualization, the y-axis in top charts does not start at zero.}
\end{figure}

These impressive results are explained by the removal of the spill bucket and the in-place partitioning procedure in LearnedSort 2.0. The removal of the spill bucket and its sorting subroutine helped LearnedSort 2.0 avoid the call to an external, comparison-based sort, resulting in significantly fewer branch mispredictions in the case of high-duplicate inputs. This was also the competitive advantage of the original LearnedSort algorithm, whose performance benefits stemmed  from the elimination of key comparisons and, thus, conditional branches. Figure \ref{fig:branches} shows that  LearnedSort 2.0 makes 9\% less branches and 75\% fewer misses as compared to the original LearnedSort in the NYC:Dist dataset (99.999\% dups). Whereas, in the case of the Standard Normal dataset (0\% dups), LearnedSort 2.0 makes 12\% fewer branches and 35\% fewer misses.

On the other hand, the use of fixed-capacity bucket fragments accounts for a smaller working set and O(1) additional memory, consequently, better cache locality for high-duplicate and large datasets. In fact, Figure \ref{fig:cache} shows that, for both Standard Normal and NYC:Dist datasets, LearnedSort 2.0 has 15\% fewer cache misses and 14\% less cache references.

These micro-benchmarks help explain the improvements of LearnedSort 2.0 that were seen from the experiments shown earlier. As a final note, we also observed dTLB and iTLB misses; however, we did not detect any significant changes in their rates for the new algorithm in either dataset. 

\section{Conclusion}
LearnedSort 2.0 is a major revision of the original LearnedSort algorithm that builds on its strongest points, while improving on the side-effects of large spill buckets on datasets containing a large portion of duplicates. This newer algorithm completely eliminates the need for the spill bucket by emulating variable-sized buckets via bucket fragments. These changes account for much better performance on high-duplicate datasets, in addition to the cases where LearnedSort was already highly optimal. 

Besides describing the new algorithm in detail, we also performed several benchmarks on a large number of datasets that cover a diverse range of distributions for various properties. We compared LearnedSort 2.0 with the original LearnedSort as well as state-of-the-art sorting algorithms and observed impressive sorting speeds. We also showed micro-benchmarks to explain its cache efficiency and the improvements from the in-place partitioning.

There still remains future work for the extension of LearnedSort on parallel execution models and the handling of disk-scale data. 

\balance

\bibliographystyle{ACM-Reference-Format}
\bibliography{bib}


\begin{thebibliography}{15}


\ifx \showCODEN    \undefined \def \showCODEN     #1{\unskip}     \fi
\ifx \showDOI      \undefined \def \showDOI       #1{#1}\fi
\ifx \showISBNx    \undefined \def \showISBNx     #1{\unskip}     \fi
\ifx \showISBNxiii \undefined \def \showISBNxiii  #1{\unskip}     \fi
\ifx \showISSN     \undefined \def \showISSN      #1{\unskip}     \fi
\ifx \showLCCN     \undefined \def \showLCCN      #1{\unskip}     \fi
\ifx \shownote     \undefined \def \shownote      #1{#1}          \fi
\ifx \showarticletitle \undefined \def \showarticletitle #1{#1}   \fi
\ifx \showURL      \undefined \def \showURL       {\relax}        \fi
\providecommand\bibfield[2]{#2}
\providecommand\bibinfo[2]{#2}
\providecommand\natexlab[1]{#1}
\providecommand\showeprint[2][]{arXiv:#2}

\bibitem[\protect\citeauthoryear{??}{nyc}{2020}]%
        {nyc}
 \bibinfo{year}{2020}\natexlab{}.
\newblock \bibinfo{title}{TLC Trip Record Data}.
\newblock
\newblock
\urldef\tempurl%
\url{https://www1.nyc.gov/site/tlc/about/tlc-trip-record-data.page}
\showURL{%
\tempurl}


\bibitem[\protect\citeauthoryear{Akira~Fujita}{Akira~Fujita}{2009}]%
        {linux_defrag}
\bibfield{author}{\bibinfo{person}{Takashi~Sato Akira~Fujita}.}
  \bibinfo{year}{2009}\natexlab{}.
\newblock \bibinfo{title}{e4defrag.c - ext4 filesystem defragmenter}.
\newblock
  \bibinfo{howpublished}{\url{https://github.com/tytso/e2fsprogs/blob/master/misc/e4defrag.c}}.
\newblock


\bibitem[\protect\citeauthoryear{Andrew~Schein}{Andrew~Schein}{2009}]%
        {radix_sort}
\bibfield{author}{\bibinfo{person}{Ani~Kristo Andrew~Schein}.}
  \bibinfo{year}{2009}\natexlab{}.
\newblock \bibinfo{title}{{Open-source C++ implementation of Radix Sort for
  double-precision floating points}}.
\newblock
\newblock
\urldef\tempurl%
\url{https://github.com/learnedsystems/LearnedSort/blob/master/third_party/radix/radix_sort.h}
\showURL{%
\tempurl}


\bibitem[\protect\citeauthoryear{Axtmann, Witt, Ferizovic, and Sanders}{Axtmann
  et~al\mbox{.}}{2021}]%
        {isp4o_benchmark}
\bibfield{author}{\bibinfo{person}{Michael Axtmann}, \bibinfo{person}{Sascha
  Witt}, \bibinfo{person}{Daniel Ferizovic}, {and} \bibinfo{person}{Peter
  Sanders}.} \bibinfo{year}{2021}\natexlab{}.
\newblock \bibinfo{title}{Engineering In-place (Shared-memory) Sorting
  Algorithms}.
\newblock
\newblock
\showeprint[arxiv]{2009.13569}~[cs.DC]


\bibitem[\protect\citeauthoryear{Blacher, Giesen, and K{\"u}hne}{Blacher
  et~al\mbox{.}}{2021}]%
        {blacher2021fast}
\bibfield{author}{\bibinfo{person}{Mark Blacher}, \bibinfo{person}{Joachim
  Giesen}, {and} \bibinfo{person}{Lars K{\"u}hne}.}
  \bibinfo{year}{2021}\natexlab{}.
\newblock \showarticletitle{Fast and Robust Vectorized In-Place Sorting of
  Primitive Types}. In \bibinfo{booktitle}{\emph{19th International Symposium
  on Experimental Algorithms (SEA 2021)}}. Schloss Dagstuhl-Leibniz-Zentrum
  f{\"u}r Informatik.
\newblock


\bibitem[\protect\citeauthoryear{Chicago}{Chicago}{2021}]%
        {chic}
\bibfield{author}{\bibinfo{person}{City~of Chicago}.}
  \bibinfo{year}{2021}\natexlab{}.
\newblock \bibinfo{title}{Taxi Trips: City of Chicago: Data Portal}.
\newblock
\newblock
\urldef\tempurl%
\url{https://data.cityofchicago.org/Transportation/Taxi-Trips/wrvz-psew#column-menu}
\showURL{%
\tempurl}


\bibitem[\protect\citeauthoryear{Edelkamp and Weiß}{Edelkamp and
  Weiß}{2016}]%
        {blockquicksort}
\bibfield{author}{\bibinfo{person}{Stefan Edelkamp} {and}
  \bibinfo{person}{Armin Weiß}.} \bibinfo{year}{2016}\natexlab{}.
\newblock \bibinfo{title}{BlockQuicksort: How Branch Mispredictions don't
  affect Quicksort}.
\newblock
\newblock
\showeprint[arxiv]{1604.06697}~[cs.DS]


\bibitem[\protect\citeauthoryear{GNU}{GNU}{2009}]%
        {std::sort}
\bibfield{author}{\bibinfo{person}{GNU}.} \bibinfo{year}{2009}\natexlab{}.
\newblock \bibinfo{title}{{C++: STL sort}}.
\newblock
\newblock
\urldef\tempurl%
\url{https://gcc.gnu.org/onlinedocs/libstdc++/libstdc++-html-USERS-4.4/a01347.html}
\showURL{%
\tempurl}


\bibitem[\protect\citeauthoryear{Inc}{Inc}{2017}]%
        {windows_defrag}
\bibfield{author}{\bibinfo{person}{Microsoft Inc}.}
  \bibinfo{year}{2017}\natexlab{}.
\newblock \bibinfo{title}{Windows defrag utility}.
\newblock
  \bibinfo{howpublished}{\url{https://docs.microsoft.com/en-us/windows-server/administration/windows-commands/defrag}}.
\newblock


\bibitem[\protect\citeauthoryear{Kipf, Marcus, van Renen, Stoian, Kemper,
  Kraska, and Neumann}{Kipf et~al\mbox{.}}{2019}]%
        {sosd}
\bibfield{author}{\bibinfo{person}{Andreas Kipf}, \bibinfo{person}{Ryan
  Marcus}, \bibinfo{person}{Alexander van Renen}, \bibinfo{person}{Mihail
  Stoian}, \bibinfo{person}{Alfons Kemper}, \bibinfo{person}{Tim Kraska}, {and}
  \bibinfo{person}{Thomas Neumann}.} \bibinfo{year}{2019}\natexlab{}.
\newblock \showarticletitle{SOSD: A benchmark for learned indexes}.
\newblock \bibinfo{journal}{\emph{arXiv preprint arXiv:1911.13014}}
  (\bibinfo{year}{2019}).
\newblock


\bibitem[\protect\citeauthoryear{Kristo}{Kristo}{2020}]%
        {bug}
\bibfield{author}{\bibinfo{person}{Ani Kristo}.}
  \bibinfo{year}{2020}\natexlab{}.
\newblock \bibinfo{title}{GitHub Issues Page}.
\newblock
\newblock
\urldef\tempurl%
\url{github.com/learnedsystems/learnedsort/issues/3}
\showURL{%
\tempurl}


\bibitem[\protect\citeauthoryear{Kristo, Vaidya, \c{C}etintemel, Misra, and
  Kraska}{Kristo et~al\mbox{.}}{2020}]%
        {kristo:sigmod2020}
\bibfield{author}{\bibinfo{person}{Ani Kristo}, \bibinfo{person}{Kapil Vaidya},
  \bibinfo{person}{Ugur \c{C}etintemel}, \bibinfo{person}{Sanchit Misra}, {and}
  \bibinfo{person}{Tim Kraska}.} \bibinfo{year}{2020}\natexlab{}.
\newblock \showarticletitle{The Case for a Learned Sorting Algorithm}. In
  \bibinfo{booktitle}{\emph{Proceedings of the 2020 ACM SIGMOD International
  Conference on Management of Data}} (Portland, OR, USA)
  \emph{(\bibinfo{series}{SIGMOD ’20})}. \bibinfo{publisher}{International
  Foundation for Autonomous Agents and Multiagent Systems},
  \bibinfo{address}{Richland, SC}, \bibinfo{pages}{1001–1016}.
\newblock
\showISBNx{9781450367356}
\urldef\tempurl%
\url{https://doi.org/10.1145/3318464.3389752}
\showDOI{\tempurl}


\bibitem[\protect\citeauthoryear{Mavrodiev}{Mavrodiev}{2019}]%
        {sof}
\bibfield{author}{\bibinfo{person}{Hristo Mavrodiev}.}
  \bibinfo{year}{2019}\natexlab{}.
\newblock \bibinfo{title}{Sofia air quality dataset}.
\newblock
\newblock
\urldef\tempurl%
\url{https://www.kaggle.com/hmavrodiev/sofia-air-quality-dataset}
\showURL{%
\tempurl}


\bibitem[\protect\citeauthoryear{Onyshchak}{Onyshchak}{2020}]%
        {stks}
\bibfield{author}{\bibinfo{person}{Oleh Onyshchak}.}
  \bibinfo{year}{2020}\natexlab{}.
\newblock \bibinfo{title}{Stock Market Dataset}.
\newblock
\newblock
\urldef\tempurl%
\url{https://www.kaggle.com/jacksoncrow/stock-market-dataset}
\showURL{%
\tempurl}


\bibitem[\protect\citeauthoryear{Skarupke}{Skarupke}{2016}]%
        {skasort}
\bibfield{author}{\bibinfo{person}{Malte Skarupke}.}
  \bibinfo{year}{2016}\natexlab{}.
\newblock
\newblock
\urldef\tempurl%
\url{https://probablydance.com/2016/12/27/i-wrote-a-faster-sorting-algorithm/}
\showURL{%
\tempurl}


\end{thebibliography}

\end{document}